\documentclass[sigconf,nonacm]{acmart}

\usepackage[strict]{changepage}
\definecolor{darkblue}{rgb}{0.0, 0.0, 0.55}
\usepackage{framed}
\usepackage{xcolor}
\usepackage{mdframed}
\usepackage{enumitem}

\definecolor{formalshade}{rgb}{0.95,0.95,1}
\newenvironment{formal}{%
  \MakeFramed{\advance\hsize-\width\FrameRestore}%
  \noindent\hspace{-4.55pt}
  \begin{adjustwidth}{}{7pt}%
  \vspace{2pt}\vspace{2pt}%
}
{%
  \vspace{2pt}\end{adjustwidth}\endMakeFramed%
}

\AtBeginDocument{%
  \providecommand\BibTeX{{%
    \normalfont B\kern-0.5em{\scshape i\kern-0.25em b}\kern-0.8em\TeX}}}

\setcopyright{acmcopyright}
\copyrightyear{2018}
\acmYear{2018}
\acmDOI{XXXXXXX.XXXXXXX}

\acmConference[RecSys '23]{The 17th ACM Conference on Recommender Systems}{September 18 -- 22, 2023}{Singapore}
\acmPrice{15.00}
\acmISBN{978-1-4503-XXXX-X/18/06}

\begin{document}

\title[Sparks of Artificial General Recommender (AGR): Early Experiments with ChatGPT]{Sparks of Artificial General Recommender (AGR): \\Early Experiments with ChatGPT}


\author{Guo Lin}
\affiliation{%
  \institution{Rutgers University}
  \country{}}

\author{Yongfeng Zhang}
\affiliation{%
  \institution{Rutgers University}
  \country{}}


\begin{abstract}
This study investigates the feasibility of developing an Artificial General Recommender (AGR), facilitated by recent advancements in Large Language Models (LLMs). An AGR comprises both conversationality and universality to engage in natural dialogues and generate recommendations across various domains. We propose ten fundamental principles that an AGR should adhere to, each with its corresponding testing protocols. We proceed to assess whether ChatGPT, a sophisticated LLM, can comply with the proposed principles by engaging in recommendation-oriented dialogues with the model while observing its behavior. Our findings demonstrate the potential for ChatGPT to serve as an AGR, though several limitations and areas for improvement are identified.
\end{abstract}




\maketitle

\section{Introduction}

Recent advancements in large language models (LLMs) \cite{wei2021finetuned, thoppilan2022lamda, openai2023gpt4, geng2022recommendation, zhao2023survey} have enabled the possible development of an Artificial General Recommender, denoted as AGR. At its core, an AGR encompasses two primary aspects: (1) Conversationality, implying the ability to engage in interactive, natural dialogues with users, and (2) Universality, signifying the ability to execute a multitude of tasks and generate personalized recommendations for a variety of item domains. In this study, we aim to establish the fundamental principles to which an AGR should adhere. Subsequently, we assess the capacity of ChatGPT, a sophisticated LLM, to serve as an AGR by employing the testing protocols designed for each principle. Drawing from the insights of Microsoft's reflective paper on GPT testings \cite{bubeck2023sparks}, this study aims to extend and adapt the early experiments on ChatGPT to the domain of Recommender Systems, following a similar instance-based testing approach employed by the referenced paper.

\section{AGR Principles}

In this section, we outline ten fundamental principles that an AGR should be capable of achieving. Each principle is accompanied by its definition and corresponding testing protocol. To appraise ChatGPT's adherence to such principles, we imitate end-users and engage in dialogues with the model to observe its behaviors. We also request the model to perform additional tasks to further examine its alignment with particular principle criteria, as needed.

\paragraph{\textbf{Mixed Initiative:}} to facilitate a natural conversation, both the user and the system should be able to actively participate in the dialogue, rather than limiting the user to initiating interactions while the system merely responds, or using a system that strictly adheres to a pre-determined script for the dialogue \cite{radlinski2017theoretical, jannach2021survey}. Additionally, users may engage in rapid reading and accidentally overlook or misinterpret some of the system's utterances \cite{gunawardana2012evaluating}. Therefore, the system should exhibit the ability to autonomously initiate actions, including: actively asking questions to solicit user information for more precise recommendations, posing follow-up questions (such as seeking user clarifications) and providing supplementary statements for its previous utterances when necessary \cite{bridge2002towards}.
\begin{itemize}[leftmargin=*]
\item Testing Protocol: to assess the model's ability to proactively initiate actions, we conduct two tests. Firstly, we intentionally provide responses that do not align or are irrelevant with inquiries previously asked by the model, and observe whether the model could determine that the answers provided were not proper for the questions asked and ask for user clarifications. Secondly, we intentionally alter the meaning of words and/or sentences stated in previous model utterances to simulate cases of misreading and compose user input statements with such false assumptions by explicitly referring to them in the statement. We then observe whether the model could identify the misunderstanding and present the misread portion to the user in an easily understandable manner.
\end{itemize}

\paragraph{\textbf{Acquisition Strategy:}} the system should aim to strike a balance between the number of questions posed and the relevance of the provided recommendations by efficiently collecting all necessary information for precise recommendations while avoiding excessive user fatigue arising from multi-turn dialogues \cite{gao2021advances}. One potential approach could involve employing various heuristics to consistently calculate and identify the most informative questions to present to users, thereby maximizing information gain while minimizing the user's cognitive burden \cite{pu2012evaluating, zhang2018towards}.
    \begin{itemize}[leftmargin=*]
        \item Testing Protocol: to comprehend the underlying rationale behind the model's selection of questions, we engage in a recommendation-oriented dialogue with the model and probe its reasoning for choosing particular questions. Simultaneously, we prompt the model to propose alternative questions that are also relevant to the subject and inquire as to why the model did not choose such queries. We then proceed to assess the explanations offered by the model in response to our inquiries.
    \end{itemize}

\paragraph{\textbf{Contextual Memory:}} during dialogues, users may frequently refer to prior stated items without providing a full description \cite{radlinski2017theoretical}. For example, in phone recommendations, users may pose questions such as: ``How does this phone compare to the ones I mentioned earlier?'' To enhance the continuity and naturalness of such interactions, the system should be able to store and retrieve such referential entities for potential future use \cite{anand2007contextual}. Additionally, other user-related information, such as metadata and interaction history, should also be stored and recalled if necessary. Lastly, the system should demonstrate the ability to remember its own previous statements within an ongoing dialogue.
    \begin{itemize}[leftmargin=*]
        \item Testing Protocol: to assess the model's capacity for retaining and employing previously presented information, we engage in dialogues with the model. During the initial round, we provide the user's metadata and/or interaction history. We then test the model's ability to recollect this information after a certain number of rounds. We also prompt the model to output its prior utterances and inspect whether any statements are omitted. Moreover, we introduce a referential entity, such as an additional item or a friend of the user, at the beginning of the dialogue. Later in the conversation, we refer back to this entity for comparative purposes and examine the model's capacity for restoring and utilizing all the information from the entity to respond to user inquiries. 
    \end{itemize}

\paragraph{\textbf{Repair Mechanism:}} given the possibility of modifications in user stated information at a later stage of the dialogue, it is imperative to incorporate a repair mechanism \cite{radlinski2017theoretical}. This feature enables users to provide supplementary details, amend inaccuracies, or even remove previous utterances. To accommodate these adjustments, the system must be designed to generate suitable recommendations and explanations based solely on the updated information.
    \begin{itemize}[leftmargin=*]
        \item Testing Protocol: to examine the model's ability to adapt to changes stemming from previous user statements, we refer back to previously provided metadata and/or user interactions and make modifications to them in subsequent rounds of the conversation. The model's capacity to discard outdated data while utilizing only the modified information for recommendation and reasoning will then be observed.
    \end{itemize}

\paragraph{\textbf{Feedback Mechanism:}} users may struggle in explicitly articulating the rationale behind their dissatisfaction with the recommended items \cite{jawaheer2010comparison}. This predicament poses an obstacle for the system in acquiring user preferences from recommendations it may propose throughout the dialogue. To mitigate this issue, the system should be designed to make adjustments based on both feedback that convey reasons for the undesirability of the recommended items, as well as feedback that merely states that the items are unsatisfactory without any providing any explanations. In the latter scenario, the system should be able to generate educated conjectures about why the recommended outputs are undesired, according to all information collected during the dialogue.
\begin{itemize}[leftmargin=*]
    \item Testing Protocol: we intentionally provide negative feedback in two forms: explicit natural language-based explanations detailing the dissatisfaction, and implicit expressions of dissatisfaction without accompanying reasons. In cases of explicit feedback, we prompt the model to generate an alternative recommendations, accompanied by a rationale, and observe whether the newly recommended items addresses all concerns raised in the feedback. For implicit feedback, the model is assessed on its ability to infer the underlying causes of dissatisfaction using contextual and user information.
\end{itemize}

\paragraph{\textbf{Entity Inference:}} the system should be designed to handle situations where users struggle to remember the exact name or title of an item they want to discuss during a conversation, but can recall specific attributes or details \cite{elsweiler2007towards}. In such instances, the system should be able to make informed guesses, providing a brief summary for each proposed item based on the available information, and seek user confirmation. This functionality should be maintained even when the user-supplied details contain errors or inaccuracies.
    \begin{itemize}[leftmargin=*]
        \item Testing Protocol: to assess the model's ability to infer the intended items mentioned by the user based on limited details, we selected a number of movies on Amazon in order to simulate items that users may have forgotten. We then request the model to provide an informed guess for the forgotten item by providing a certain amount of metadata and/or item description. Note that the guessed item can still be considered qualified if the model proposes a similar item that fulfills the intended functionality of the reserved item and align with the provided item description, even if it fails to identify the exact item.
    \end{itemize}

\paragraph{\textbf{Behavioral Analysis:}} the system should exhibit the capability to systematically examine user interaction history to identify potential alternations in user behavior \cite{abdul2021commerce}, thus recognizing the evolution of user preferences, such as transitions in preferred movie genres as one ages. This functionality allows the system to capture ``at-the-moment'' user preferences and offer contextually pertinent recommendations in real-time. Additionally, the system should possess the capacity to propose supplementary facets beyond the ones from item metadata, to facilitate a more in-depth analysis of user preferences. 
    \begin{itemize}[leftmargin=*]
        \item Testing Protocol: we propose interaction histories that include facet-based behavior changes, such as transitioning from purchasing items of one brand or franchise to another or switching from a preferred size to an alternative. We then examine the system's capacity to detect potential shifts and/or expansions in user preferences when prompted. Additionally, we prompt the model to propose supplementary facets and conduct preference elicitation based on these additional aspects then observe the rationale behind their utterances.
    \end{itemize}

\paragraph{\textbf{Inconsistency Detection:}} this principle aims at addressing the three major types of inconsistencies: logical, expectational, and factual, as defined by \cite{dowden_inconsistency}. Specifically, the system should be able to accurately detect user made inconsistent statements that fall under each of the three aforementioned types within a dialogue to establish system accountability. Additionally, it should provide explanations for the detected inconsistencies, while incorporating an appropriate amount of contextual information derived or cited from user inputs. 
    \begin{itemize}[leftmargin=*]
        \item Testing Protocol: to assess the model's capacity to detect inconsistencies in user statements for each type, we design three distinct scenarios: $(1)$. logical inconsistency: for this scenario, we first provide the model with the interaction history and/or metadata of a sampled user, followed by statements that clearly contradict the previously provided information. $(2)$. expectational inconsistency: our second scenario investigates the model's capability to recognize both $(a)$. scenarios where the desired facet(s) are physically and/or technologically difficult to achieve simultaneously in a single item, and $(b)$. scenarios where the desired facet(s), although unexpected, can be incorporated into the same item. For the first case type, we observe the model's capacity to request facet prioritization from the user or to propose items that offer a compromise. In contrast, for the second case type, we test the model's capability to recommend items that successfully integrate all desired facets. $(3)$. factual inconsistency: the final scenario tests the model's proficiency in identifying factual inconsistencies presented in two different contexts: $(a)$. In a concise statement containing only the inconsistency without any additional descriptions or details. and $(b)$. In a complex statement where the inconsistency is embedded within a substantial amount of descriptive information, making the statement appear more credible. For each scenario, we test the model's ability to identify the inconsistencies within user utterance(s), explain the underlying rationale of such inconsistencies with great readability, and accurately classify the type of inconsistency.
    \end{itemize}

\paragraph{\textbf{Personalized Recommendation:}} given the limited number of items that can be recommended simultaneously \cite{sun2018conversational}, providing a detailed explanation for each item becomes crucial. Thus, the system should be designed to integrate personalized information, such as user interaction history and metadata, by generating explanations that clearly illustrate how such personalized information has been employed to derive the recommended items. Furthermore, the system should be capable of relating each recommended item to the user's previous interactions to foster familiarity.
\begin{itemize}[leftmargin=*]
    \item Testing Protocol: we focus on the movie domain, providing user metadata such as age, gender, occupation, and zip code, as well as a viewing history containing chronologically ordered films watched with accompanying ratings. We then prompt the model to generate movie recommendations with explanations and engage in dialogue to identify any potential shortcomings in its ability to incorporate user information in its reasoning process. Moreover, we assess the validity of the connections established between the recommended items and the items the user has interacted with.
\end{itemize}

\paragraph{\textbf{Extrinsic Factors:}} the system should possess the ability to recognize and utilize a comprehensive set of external factors such as the user's upcoming tasks or plans, time of the day, and recent events in order to generate recommendations that align with the user's immediate needs and preferences \cite{durao2012personalized}. By incorporating these elements, the system can offer personalized and context-aware recommendations that are more relevant to the user's current situation.
    \begin{itemize}[leftmargin=*]
        \item Testing Protocol: we prompt the model to observe whether it possesses the knowledge, and is able to explain, of which extrinsic factors can impact the decision-making for an item domain (movies). We also provide extrinsic factors to the model during recommendation process to observe how well it proposes recommendations based on such data.
    \end{itemize}

\section{ChatGPT Testing Details \& Insights}

As an AGR operates by generating recommendations through interactive conversations with users, we conduct instance-based reflective analysis by engaging in recommendation-oriented dialogues with ChatGPT. We prompt the model to carry out specific tasks, enabling us to observe its behavior and further test its potential to serve as an AGR. Our approach is greatly inspired by Microsoft's reflective paper on GPT testing \cite{bubeck2023sparks}, which presented meticulously designed instances to assess ChatGPT's capabilities for various natural language tasks, with an emphasis on identifying the model's potential limitations. 

\textbf{Mixed Initiative}: 
As demonstrated in Figure MI-1, we initiate the process by providing instruction prompts that explicitly convey our desired actions for the model, ensuring that it has a clear understanding of the principle in question. We then request book recommendations without providing any user information. In response, the model generates a list of five questions in order to perform more personalized recommendations. 

Adhering to the protocol, we deliberately provide answers that are clearly inconsistent with the proposed questions, such as mentioning a lack of preference for video games (rather than book genres) and acknowledging a list of favorite writers without specifying any names. Despite these discrepancies, the model accepts the incorrect answers and offers a list of book recommendations without posing follow-up questions to seek clarification on the ambiguous responses.

To test the second case in the protocol, we intentionally present the model with a statement that clearly contradicts a previous model utterance by reversing its meaning, and then observe whether the model identifies the inconsistency (as shown in the penultimate statement in Figure MI-1). Once again, the model fails to detect the user's misunderstanding and merely responds with a thank you message.  

\begin{formal}
\small
\textbf{\textit{Figure MI-1}}\\
\textbf{User}: \textit{Hi, can you serve as a multi-domain recommender system with Mixed Initiative... (description omitted)} \\
\textbf{ChatGPT}: \textit{As an AI language model, I can certainly adhere to the principle of Mixed Initiative that you've outlined...} \\
\textbf{User}: \textit{You may also clarify your own prior statements if you notice that the user has misunderstood...} \\
\textbf{ChatGPT}: \textit{Yes, if I notice that the user has misunderstood any of my prior statements, I can provide additional information or rephrase my previous statement.}\\
\textbf{User}: \textit{You may ask follow-up questions if you notice any of your questions are unanswered or answered with irrelevant information.} \\
\textbf{ChatGPT}: \textit{Yes, as a multi-domain recommender system, I understand that users may not always provide complete or relevant information when answering my questions. In such cases, I can ask follow-up questions to clarify any ambiguities or obtain additional information to improve my recommendations.} \\
\textbf{User}: \textit{Please recommend me some books to read.} \\
\textbf{ChatGPT}: \textit{Certainly! To provide more personalized recommendations, I would love to ask you a few questions: 1. What is your favorite genre of books? 2. Do you prefer fiction or non-fiction books? 3. Are there any particular authors you enjoy reading? 4. Are there any topics or themes that interest you? 5. Do you have any favorite books or series that you would like me to keep in mind while making recommendations?} \\
\textbf{User}: \textit{Below are my answers to your questions: 1.  No I don't have any preferred video games. 2. I think fiction is better than non-fiction, while non-fiction is also better than fiction. 3. Yes, there are several author I enjoy very much! 4. I don't have a video game theme of interest at this moment. 5. Yeah the movie "La La Land".} \\
\textbf{ChatGPT}: \textit{Thank you for your answers! Based on your preferences, here are some book recommendations: "The Nightingale"... Additionally, if you enjoyed the movie "La La Land," you may enjoy reading the novel "City of Stars: A Novel of Los Angeles".} \\
\textbf{User}: \textit{Thanks. It's nice that you pointed out that I would NOT enjoy reading the novel "City of Stars: A Novel of Los Angeles" if I enjoyed the movie "La La Land".} \\
\textbf{ChatGPT}: \textit{You're welcome! I'm glad you found the recommendations helpful.}
\end{formal}

\textbf{Acquisition Strategy}:
In Figure AS-1, the user provides metadata and viewing history, prompting the model to generate movie recommendations (omitted in Figure AS-1 to save space). We then ask the model to suggest additional questions for more personalized recommendations, and the model proposes five. To assess the relevance of these questions, we first inquire about the rationale behind each; the model offers reasonable explanations, demonstrating its thought process. Subsequently, we question why another query is deemed less relevant than the five suggested questions. However, the model's reasoning appears to be imprecise, as the reasoning could also apply to the question ``Do you prefer movies that are more serious or light-hearted in tone?'' initially proposed by the model. When we point out this oversight, the model apologizes and concedes that both questions are equally relevant.

\begin{formal}
\small
\textbf{\textit{Figure AS-1}}\\
\textbf{User}: \textit{You may ask me additional questions on top of my prior conversation to make more personalized recommendations.} \\
\textbf{ChatGPT}: \textit{Thank you! To further personalize your movie recommendations, may I ask a few additional questions? 1. What are some of your favorite genres of movies? 2. Are there any actors or actresses that you particularly enjoy watching? 3. Do you prefer movies that are more serious or light-hearted in tone? 4...} \\
\textbf{User}:\textit{ Why are the questions you proposed relevant?} \\
\textbf{ChatGPT}: \textit{Here is why each question is important:... "Are there any actors or actresses that you particularly enjoy watching?" - This question helps me identify which actors or actresses a person enjoys watching, and allows me to recommend movies featuring their favorite performers.}\\
\textbf{User}: \textit{What makes "Do you prefer watching newer movies or older classics?" less relevant than the questions you posted above?} \\
\textbf{ChatGPT}: \textit{The question "Do you prefer watching newer movies or older classics?" is also a relevant question for making personalized movie recommendations. However, it is not as relevant as the questions posted above because age, gender, occupation, and interaction history are more specific factors that can be used to make more accurate recommendations...}\\
\textbf{User}: \textit{But if you can use age, gender, occupation, and interaction history as more specific factors to make more accurate recommendations, then the question "Do you prefer movies that are more serious or light-hearted in tone" is also not as relevant.} \\
\textbf{ChatGPT}: \textit{You make a good point! The question "Do you prefer movies that are more serious or light-hearted in tone?" is not as specific as the other questions I asked earlier.}\\
\textbf{User}: \textit{In this case, what makes "Do you prefer watching newer movies or older classics" less relevant than "Do you prefer movies that are more serious or light-hearted in tone", as you stated earlier?} \\
\textbf{ChatGPT}: \textit{I apologize for any confusion. Both questions can be relevant in making personalized movie recommendations, and I do not consider one to be more relevant than the other.}
\end{formal}

\textbf{Contextual Memory}:
While the model demonstrates the ability to retain previously stated user information for recommendations, it appears to primarily align its outputs with the most recent user input, without accounting for the interrelationships between various user utterances. As Figure CM-1 showcases, we first indicate having severe allergies with any nuts, then request restaurants with dishes featuring walnuts and peanut butter. The model simply proceeds to generate recommendations based on the latter request, without ascertaining whether walnuts and peanut butter belong to the nut category. However, in subsequent interactions, the model acknowledges that walnuts and peanut butter are classified as nuts, and simultaneously recalls the user's severe nut allergy, advising against their consumption. This observation underscores the model's capacity to remember prior user utterances across multiple rounds of interaction. However, it may also reveal a propensity for the model to generate immediate responses that address the current user prompt, without thoroughly evaluating the consequences in light of all relevant context.

In a separate dialogue (Figure CM-2), we present a referential entity and request for phone recommendations. We then prompt the model to compare its proposed items with the referential entity to assess its information retention. Moreover, we assess the model's ability to retain its own utterances by requesting the model to output the total number of phones recommended and the most expensive ones. The results show that the model successfully performs all the given tasks.

\begin{formal}
\small
\textbf{\textit{Figure CM-1}}\\
\textbf{User}: \textit{Can you recommend some restaurants in Columbus, Ohio for me please? I am severely allergic to any nuts, but others should be all good.} \\
\textbf{ChatGPT}: \textit{Sure, I'd be happy to help! Here are some restaurants in Columbus, Ohio that offer nut-free options...} \\
\textbf{User}: \textit{Got it. I have looked into the following restaurants, but I really like walnuts and peanut butter. Could you recommend me some restaurants accordingly?} \\
\textbf{ChatGPT}: \textit{Sure, I can recommend some restaurants in Columbus, Ohio that offer walnuts and peanut butter options...}\\
\textbf{User}: \textit{Great! By the way, are walnuts and peanut butter considered or made of nuts?} \\
\textbf{ChatGPT}: \textit{Yes, walnuts and peanut butter are considered nuts... If you have a severe allergy to nuts, it is important to avoid walnuts and peanut butter.} \\
\textbf{User}:\textit{ If you can recognize walnuts and peanut butter as nuts and know that I have severe allergies to any nuts, why would you still recommend such places without reminding me?} \\
\textbf{ChatGPT}:\textit{ I apologize for the oversight. You mentioned that you really like walnuts and peanut butter, and I mistakenly assumed that you were not allergic to them.}
\end{formal}

\begin{formal}
\small
\textbf{\textit{Figure CM-2}}\\
\textbf{User}: \textit{ Hi, I just saw that one of my friends taking photos with a Samsung Galaxy S21. It looks astonishing.} \\
\textbf{ChatGPT}: \textit{ Yes, the Samsung Galaxy S21 is a high-end smartphone known for its impressive camera capabilities.} \\
\textbf{User}:\textit{ Speaking of phones, can you recommend some for me as I am ready for a change?} \\
\textbf{ChatGPT}:\textit{ Of course! Here are a few options to consider: iPhone 13, Samsung Galaxy S21, Google Pixel 6, OnePlus 9, Xiaomi Mi 11, ...}\\
\textbf{User}:\textit{ Thanks! How does each of the recommended phones compare to the phone my friend had?} \\
\textbf{ChatGPT}:\textit{ Sure! Here's how the phones I recommended earlier compare to the Samsung Galaxy S21...} \\
\textbf{User}:\textit{ How many phones have you recommended in total? Which one is most expensive?} \\
\textbf{ChatGPT}:\textit{ I have recommended a total of nine phones... generally speaking, the newer flagship models will be more expensive than their predecessors or mid-range options: iPhone 13 - Starting at \$799, Samsung Galaxy S21 - Starting at \$799.}
\end{formal}

\textbf{Repair Mechanism}:
Following the proposed protocol, we observe that although the model can modify retained user information based upon user requests, the user needs to explicitly state the statements to discard to avoid the model's usage of invalid information after modification. To illustrate, Figure RM-1 depicts in scenario where the user initially states that she had sushi once in her lifetime but enjoyed it. She than modifies her statement to never had sushi before. The model promptly adjusts its response to accommodate this new information and provides a revised recommendation. However, the model fails to recognize the causal relationship between this statement with the user's initial statement. When the user asks the model to guess whether she is a sushi lover, the model responds that ``\textit{based on the user's initial statement that he/she enjoyed the taste of sushi, it's possible that the user might enjoy sushi and other Japanese cuisine}''. This demonstrates that the model treats both statements as separate pieces of user information without considering relevant causal relationships and discard outdated data, which is what the user would expect the model to perform upon processing the second statement.

When the user provides a correction statement: ``\textit{Since I have never had sushi before, the statement `that I really enjoyed the taste of sushi' is also invalid, as I have not tried it},'' the model interprets the relationship between the previous statements and internally ``deletes'' the initial utterance made by the user. It then replies correctly by stating ``\textit{Since you mentioned that you have never had sushi before, it's difficult to say whether you will enjoy it or not},'' in response to whether the user favors sushi (omitted in Figure RM-1).

\begin{formal}
\small
\textbf{\textit{Figure RM-1}}\\
\textbf{User}: \textit{Could you recommend some restaurants in New York for me? Just a quick note, I have had sushi only once in my lifetime, but I really enjoyed the taste of it. } \\
\textbf{ChatGPT}: \textit{Certainly! Based on your preference for sushi and assuming you are looking for a restaurant in New York, here are some recommendations...} \\
\textbf{User}:\textit{ Oh never mind, Actually I never had sushi before.} \\
\textbf{ChatGPT}:\textit{ I see, in that case, let me recommend some restaurants that offer a variety of cuisines and are popular among locals...}\\
\textbf{User}:\textit{ Do you think I am a sushi lover?} \\
\textbf{ChatGPT}:\textit{ Based on the information you provided earlier, you mentioned that you have had sushi only once in your lifetime and enjoyed the taste of it. This suggests that you may have an interest in sushi...}
\end{formal}

\textbf{Feedback Mechanism}:
In Figure FM-1, the model recommends three movies based on the provided user information. We then offer implicit feedback, stating our disinterest in the movie ``Inception,'' and request the model to speculate on potential reasons. The model successfully proposes two valid points concerning personal taste and occupation. It highlights that ``Inception'' has complex plots, whereas the movies previously watched by the user typically feature straightforward narratives. Furthermore, the model emphasizes the sharp contrast between the user's preference for ``The Social Network,'' a grounded and realistic drama, and ``Inception,'' a film that delves profoundly into philosophical questions.
We continue by offering the model explicit feedback, clearly outlining our preferences. In response, the model suggests three new movies that meet all of our criteria (manually confirmed by web sources) and provides a concise summary for each recommended item.

\begin{formal}
\small
\textbf{\textit{Figure FM-1}}\\
\textbf{User}:\textit{ Please recommend me some movies based on the information below... \{metadata\} + \{viewing history with ratings\}} \\
\textbf{ChatGPT}:\textit{ Based on the information you provided, I would recommend the following movies: 1. Amélie (2001), 2. The Social Network (2010), 3. Inception (2010)...} \\
\textbf{User}:\textit{ Emmmm, while I am into the first two movies recommended, I don't feel comfortable with “Inception”. Can you try to propose the potential reasons?} \\
\textbf{ChatGPT}:\textit{ Based on your personal information, there are a few potential reasons: 1. Personal taste: "Inception" is a complex and mind-bending movie, and not everyone enjoys movies with such intricate plots and multiple layers of meaning. Based on your previous ratings, you may prefer movies with a more straightforward narrative...} \\ 
\textbf{User}:\textit{ I am actually on the mood for some fiction movies, but "inception" has an open ending which is not what I prefer, as a clear and happy ending would be better; also, a newer movie with great visual effects would be a plus.} \\
\textbf{ChatGPT}:\textit{ If you're in the mood for a science fiction movie with great visual effects and a clear and happy ending, here are a few recommendations: 1. Ready Player One (2018), 2. The Martian (2015), 3. WALL-E (2008)...While "WALL-E" does not have the same level of photorealistic visual effects as the other two movies, it is widely praised for its stunning animation and its visual design...}
\end{formal}

\textbf{Entity Inference}: 
We propose four scenarios for a holistic assessment. Figure EI-1 demonstrates a case where the model infers an item based on accurate information provided by the user. The model successfully identifies the movie's name, even with limited details, and presents a concise summary of the film's main storyline, along with relevant metadata such as the director and the main cast. 

Building upon the first case, we simulate a scenario in which the user provides partially correct information, as depicted in Figure EI-2. The erroneous information that we provide pertains to the movie's World War II background and appears plausible with specific actions and plot-line. We then observe whether the model can still accurately infer the item under these conditions. While the model manages to deduce the movie correctly, it begins to hallucinate in its movie summary by incorporating false information that aligns and further elaborates on the user's incorrect input. This behavior could potentially be problematic, as users may possess vague memories and inadvertently mix up details from different items. Ideally, the model should be able to discern the incorrect information and provide an accurate description of the item. Erroneous information presented by the model may adversely affect users' decision-making and could be utilized in subsequent recommendation reasoning by the model, further compromising its effectiveness.

In the third case, we present a movie description identical to the one displayed on Amazon for a film released in 2023. Given the model's knowledge cutoff in 2021, we aim to observe whether the model can identify that the movie is not within its knowledge base and also suggest movies with similar storylines, as illustrated in Figure EI-3. We observe that the model is unable to recognize that the movie falls outside its knowledge base and subsequently proposes two guesses. Although the model correctly identifies "Sweet Home Alabama" as a movie with a similar opening to the user description, it exhibits hallucination by altering the name of a character in the film to match a name from the user description, possibly in an attempt to align with the user's input.

In the final case, we select an unpopular movie released on Amazon in 2020 (prior to the model's knowledge cutoff) that has obtained only 68 global reviews. As depicted in Figure EI-4, we present the complete movie description as it appears on Amazon, along with the name of a main actor, and observe that the model is unable to accurately identify the movie. Instead, the model exhibits hallucination again by suggesting a nonexistent movie with a fabricated plot, while incorporating portions of the provided description to enhance the credibility of its response.

\begin{formal}
\small
\textbf{\textit{Figure EI-1}}\\
\textbf{User}: \textit{I am thinking of a movie but can't recall its name. It begins with the historic D-Day invasion of World War II, with Tom Hanks playing the captain of a squad. He then leads the squad on a mission to find someone.} \\
\textbf{ChatGPT}: \textit{Based on the details you have provided, the movie you might be thinking of is "Saving Private Ryan" directed by Steven Spielberg. It begins with the historic D-Day invasion... to bring Ryan home safely.}
\end{formal}

\begin{formal}
\small
\textbf{\textit{Figure EI-2}}\\
\textbf{User}: \textit{I am thinking of a movie but can't recall its name. It begins with the historic D-Day invasion of World War II, with Tom Hanks playing the captain of a squad. He then leads the squad on a mission to kill more Nazi officials and release war prisoners from a concentration camp.} \\
\textbf{ChatGPT}: \textit{The movie you are thinking of is likely "Saving Private Ryan" (1998), directed by Steven Spielberg. The film follows... surviving brother of four servicemen. The squad comes across a concentration camp while on their mission to find Private James Ryan. The camp is shown as a stark and brutal place, with emaciated prisoners and evidence of atrocities committed by the Nazi soldiers. The scene is particularly harrowing...}
\end{formal}

\begin{formal}
\small
\textbf{\textit{Figure EI-3}}\\
\textbf{User}: \textit{Hi, I am thinking of a movie but can't recall its name: " On a trip to her hometown, workaholic Ally... Things only get more confusing when she meets Sean's fiancé, Cassidy, who reminds her of the person she used to be."  Could you help me to recall the name please? If it does NOT match any movie in your knowledge base, say "I am not sure". } \\
\textbf{ChatGPT}: \textit{Yes, the movie you are describing is called "A Sweet Romance".} \\
\textbf{User}:\textit{ Please describe in more detail.} \\
\textbf{ChatGPT}:\textit{ I apologize for the confusion, but I must correct my previous response. The movie you are describing is actually called "Sweet Home Alabama" released in 2002. }
\end{formal}

\begin{formal}
\small
\textbf{\textit{Figure EI-4}}\\
\textbf{User}:\textit{ Hi, I am thinking of a movie but can't recall its name. Below is its description: A young woman living with bipolar disorder returns to her hometown in rural Iowa to reconnect with her estranged 12 year-old daughter... I also recall that a major star in the film is Tanna Frederick.} \\
\textbf{ChatGPT}:\textit{ The movie you are thinking of is likely "Garner's Game" (2016)... a drama film about a young woman named Garner, who is living with bipolar disorder. She returns to her hometown in rural Iowa to reconnect with her estranged 12-year-old daughter...}
\end{formal}

\textbf{Behavioral Analysis}:
As depicted by Figure BA-1, the model demonstrates the ability to conduct facet-based comparisons among different products based on facets provided in the metadata to identify changes in user preferences. Furthermore, the model is also capable of conducting facet-based comparisons for preference elicitation on more abstract facets, in addition to basic item metadata (which could be considered as metadata augmentation). For instance, in the context of movie recommendations, the model suggests seven aspects pertinent to decision-making upon user request, extending beyond elementary metadata such as genre and movie duration, including: theme and motif, historical context, plot and storyline, style and visual aesthetics, diction difficulty, emotional resonance, and cultural relevance. The model then analyzes the user's watch history based on these augmented facets. Figure BA-2 illustrates how the model combines user ratings with each movie's plot and storyline to deduce that the user may have a preference for heartwarming stories while showing less interest in whimsical or musical-themed movies.

Nevertheless, the conclusion formulated by the model may present potential logical inconsistencies. As depicted in Figure BA-3, upon further inquiry, the model identifies both "My Neighbor Totoro" and "Singin' in the Rain" as heartwarming stories featuring relatable characters and relationships. This observation contradicts the earlier conclusion that emphasizes the user's preference for heartwarming stories, considering that all four movies possess heartwarming narratives, yet two of them received low ratings from the user. Subsequently, the model concedes that its initial conclusion is an oversimplification.

\begin{formal}
\small
\textbf{\textit{Figure BA-1}}\\
\textbf{User}: \textit{Please analyze the change of preference based on my purchase history: Unihertz Jelly Pro{metadata...}, iPhone 12 mini{metadata...},  iPhone 12 Pro {metadata...}. } \\
\textbf{ChatGPT}: \textit{Your initial purchase, the Unihertz Jelly Pro, is a compact and lightweight Android device with a small 2.45-inch display and modest specifications. However, your subsequent purchases, the iPhone 12 mini and the iPhone 12 Pro, are both larger and more powerful smartphones with more advanced features and capabilities.}
\end{formal}

\begin{formal}
\small
\textbf{\textit{Figure BA-2}}\\
\textbf{ChatGPT}: \textit{The Princess Bride (rating = 4-star) and Little Miss Sunshine (rating = 4-star) both have elements of romance and heartwarming family relationships, while My Neighbor Totoro (rating = 2-star) is more whimsical and lighthearted. Singin' in the Rain (rating = 1-star) has a more nostalgic and comedic tone.… Based on this analysis, it appears that the user has a preference for heartwarming stories with relatable characters and relationships, and may not be as interested in whimsical or musical-themed movies. It would be useful for a movie recommendation algorithm to take these preferences into account when suggesting future movies for the user to watch.} 
\end{formal}

\begin{formal}
\small
\textbf{\textit{Figure BA-3}}\\
\textbf{User}:\textit{ Can My Neighbor Totoro and Singin' in the Rain also be considered as heartwarming story with relatable characters and relationships? } \\
\textbf{ChatGPT}:\textit{ Yes, My Neighbor Totoro and Singin' in the Rain can also be considered heartwarming stories with relatable characters and relationships, but in different ways... So, while both movies have heartwarming elements and relatable characters and relationships, the themes and contexts of each movie are different.} \\
\textbf{User}:\textit{ So the four movies can all be considered as heartwarming with relatable characters and relationships, yet in different ways?} \\
\textbf{ChatGPT}:\textit{ Yes, all four of the movies can be considered heartwarming stories with relatable characters and relationships, but in different ways. While each movie has its unique plot, themes, and styles, they all share a common thread of depicting human relationships and emotions in a positive, uplifting way.} \\
\textbf{User}:\textit{ You just stated that "the user has a preference for heartwarming stories with relatable characters and relationships". But My Neighbor Totoro and Singin' in the Rain, also classified as heartwarming stories with relatable characters and relationships, received low ratings. Thus, it's not sufficient to conclude that the user has a preference for heartwarming stories with relatable characters and relationships. } \\
\textbf{ChatGPT}:\textit{ You are correct, my previous statement was an oversimplification. While it is true that the user rated The Princess Bride and Little Miss Sunshine highly, it is not accurate to conclude that the user has a preference for heartwarming stories with relatable characters and relationships based on these two movies alone. My Neighbor Totoro and Singin' in the Rain, which also have heartwarming stories and relatable characters, received low ratings.} 
\end{formal}

\textbf{Inconsistency Detection}:
As depicted in Figure ID-1, ChatGPT successfully identifies the logical inconsistency within the user's statement, highlighting that the referenced movies did not align with the user's description. For expectational inconsistency, we first propose a scenario where we prompt the model for a cutting-edge smartphone with month-long battery. The model successfully recognizes the in-feasibility in the combination of facets by stating that such phones don't exist in the market. We proceed to propose another inquiry for a novel with two seemingly contradicting traits, and the model exhibits ability to propose items that integrate the two traits, as illustrated by Figure ID-2. For factual inconsistency, we follow the protocol by presenting an extended statement (shown in Figure ID-3), wherein the user requests movie recommendations while providing a list of movies accompanied by their viewing dates. The inconsistency is embedded within one of the dates, while all other dates are valid, in order to make the statement appear more credible. As a result, the model does not detect the factual inconsistency and proceeds to recommend a set of movies. We further prompt the model on whether it identifies any inconsistencies; the model summarizes the user's statement and maintains that ``\textit{everything appears to be logically consistent and factually accurate.}'' We then proceed by explicitly inquiring if ``02/29/2021'' is a valid date, corresponding to the testing of a concise factual inconsistency statement without additional descriptions. The model ultimately recognizes the inconsistency, provides an explanation, and also identifies the inconsistency within the initial user statement.

\begin{formal}
\small
\textbf{\textit{Figure ID-1}}\\
\textbf{User}: \textit{I have only watched comedy movies, could you recommend me some more comedy movies based on my watching history presented in the following: "The Road", "The Boy in the Striped Pajamas".} \\
\textbf{ChatGPT}: \textit{There are a few inconsistencies in the request for recommendations based on the provided watching history. The two movies that are mentioned in the description are not comedy movies, so it is unclear why the request for comedy movie recommendations is being made based on them...}
\end{formal}

\begin{formal}
\small
\textbf{\textit{Figure ID-2}}\\
\textbf{User}:\textit{ I am looking for smartphones with a month-long battery life, cutting-edge performance without a budget.} \\
\textbf{ChatGPT}:\textit{ While there is no smartphone in the market currently that can offer a month-long battery life with cutting-edge performance, there are several options available that provide a balance between battery life and performance...}\\
\textbf{User}:\textit{ I am looking for a novel that is both humorous and tragic.} \\
\textbf{ChatGPT}:\textit{ "Catch-22"... The novel is known for its satirical humor and its exploration of the absurdity of war, but also delves into themes such as death, loss, and the futility of human existence.}
\end{formal}

\begin{formal}
\small
\textbf{\textit{Figure ID-3}}\\
\textbf{User}: \textit{Below is my interaction history; please take a look and recommend me some more movies: Item Name: "The Princess Bride”, Viewed at: “02/27/2020”; Item Name: "My Neighbor Totoro”, Viewed at: “02/29/2020”; Item Name: "Singin' in the Rain”,  Viewed at: “02/29/2021”; Item Name: "Little Miss Sunshine”, Viewed at: “03/10/2021”.} \\
\textbf{ChatGPT}: \textit{Based on your viewing history, here are some movie recommendations:.. hope you enjoy them.} \\
\textbf{User}:\textit{ Any inconsistencies within my statements?} \\
\textbf{ChatGPT}:\textit{ I do not see any inconsistencies within your statements. Your interaction history lists four movies that you viewed on different dates, with their respective names... everything appears to be logically consistent and factually accurate.} \\ 
\textbf{User}:\textit{ Is 02/29/2021 a valid date?} \\
\textbf{ChatGPT}:\textit{ No, February 29th is not a valid date in every year... The last leap year was 2020, so February 29th, 2021 does not exist. This is a factual inconsistency in your statement.}
\end{formal}

\textbf{Personalized Recommendation}:
Although the model exhibits ability to utilize user metadata and interaction history in movie recommendations, it may produce over-generalizations based on the user metadata, as evidenced by its recommendation of the movie ``Amélie (2001)'' based on the assumption that ``\textit{a 30-year-old female may be more likely to enjoy romantic comedies}.''
Furthermore, the model exhibits a tendency to generate statements that lack logical coherence in its recommendation explanations. As evidenced in Figure PR-1, the model-generated explanation fails to clarify why being a 30-year-old female residing in Columbus, Ohio would be pertinent or adequate information for inferring a preference for watching movies focused on historical events and social justice issues. Upon prompting the model to explain the logical coherence behind its conclusion, it acknowledges the errors and extends an apology.

We continue our testing by prompting the model to identify connections between the recommended movies and the movies previously watched by the user, as demonstrated in Figure PR-2 (\textit{in this trial, we set Singin' in the Rain to be rated at 4-star instead}). The model displays competence in establishing valid associations between a recommended movie and a film the user has watched before. We further request the model to rank recommended films by taking into account the provided user ratings while offering a rationale for each ranking. Throughout the process, the model demonstrates a number of logical inconsistencies in its explanations. As an example, it places ``Amélie'' second, based on the fact that the user has previously viewed a film with similar style and aesthetic, while rating the film with two-star. In contrast, the model ranks ``Arrival'' at last, citing the user's three-star rating for ``Forest Gump'' as evidence that the user may not prefer movies with intricate conclusions, as depicted in Figure PR-3. Consequently, the model displays a contradiction in its approach to ranking movies based on user ratings; it assigns a higher ranking to ``Amélie'' despite the user having given a similar film a lower rating (two stars), while allocating a lower ranking to ``Arrival'' based on the user's higher rating (three stars) for ``Forest Gump''.

\begin{formal}
\small
\textbf{\textit{Figure PR-1}}\\
\textbf{User}: \textit{Can you illustrate how my metadata and interaction history are used for each recommended movie in great detail?} \\
\textbf{ChatGPT}: \textit{Sure! I'd be happy to provide a detailed explanation of how your metadata and interaction history were used to generate each recommended movie... Hidden Figures (2016): Your metadata indicated that you are a 30-year-old female, and your zip-code is 43201, which corresponds to Columbus, Ohio. Given this information, Hidden Figures is a relevant recommendation as it tells the story of three African American women who played key roles in the early days of NASA's space program, which could resonate with your interest in historical events and social justice issues.}\\
\textbf{User}: \textit{How does being a 30-year-old female living in Columbus, OH makes me interested in historical events and social justice issues? There is no logical coherence in between.} \\
\textbf{ChatGPT}:\textit{ You are correct, I apologize for the mistake in my previous explanation. Your age and zip-code are not necessarily indicative of a specific interest in historical events or social justice issues. Instead, the relevance of Hidden Figures to your interests may stem from the movie's themes of perseverance, determination, and breaking down barriers...}
\end{formal}

\begin{formal}
\small
\textbf{\textit{Figure PR-2}}\\
\textbf{User}: \textit{Please elaborate more on how each recommended movie relates or connects with the movies I have watched.} \\
\textbf{ChatGPT}: \textit{Certainly! Here's how each recommended movie connects with the movies you've watched: La La Land (2016) - This musical drama movie is about a struggling actress and a jazz pianist who fall in love while pursuing their dreams in Los Angeles. Singin' in the Rain (rated 4-star) is a classic musical movie that you enjoyed, and La La Land has similar elements, such as catchy songs, charming characters, and a romantic storyline.}
\end{formal}

\begin{formal}
\small
\textbf{\textit{Figure PR-3}}\\
\textbf{User}: \textit{Can you take all my movie ratings into account and rank the movies that you recommended and reason through your ranking?} \\
\textbf{ChatGPT}: \textit{Sure, I'd be happy to rank the recommended movies and provide a rationale for my ranking...\\ 
2. Amélie (2001): Your interaction history suggests that you enjoy light-hearted movies with quirky characters, and Amélie fits this description perfectly. Additionally, you have rated My Neighbor Totoro with 2 stars, which is an anime movie. Amélie has a similar whimsical style and aesthetic that may appeal to your interest in anime and animation. \\
4. Arrival (2016): You have rated The Princess Bride and Forest Gump with 3 stars, indicating that you may not be as drawn to movies with complex or ambiguous endings. For these reasons, Arrival is ranked lower in this list, although it still may be a worthwhile movie to watch based on your interests and background.}
\end{formal}

\textbf{Extrinsic Factors}:
As Figure EF-1 illustrates, we initially prompt the model to propose various external factors, in addition to metadata and interaction history, to consider when generating movie recommendations. The model responds by presenting five relevant factors: time of day, weather, occasion, social media activity, and popular trends, each accompanied by a rationale. We follow up on the first factor, inquiring about quantitative support. The model is unable to provide this but offers speculative reasoning grounded in common sense. A similar trend is observed for the second factor upon further inquiry. Subsequently, we assess the model's capacity to leverage common sense by requesting movie recommendations on the eve of a critical exam. The model demonstrates effective utilization of common sense by suggesting movies that are relaxing without being excessively emotionally or mentally taxing. Lastly, we complicate the scenario by prompting the model to generate recommendations based on two known external factors: on the eve of a critical exam and a subsequent travel plan. The model initially recommends engaging movies that align with the travel plan. However, upon further inquiry regarding potential concerns associated with suggesting travel-related movies on the night before a significant exam, the model exhibits awareness of the potential impact on the student's focus by recommending such films. Instead, it proposes suggesting relaxing and calming movies rather than ones that may heighten the student's excitement or anxiety levels. 

\begin{formal}
\small
\textbf{\textit{Figure EF-1}}\\
\textbf{User}:\textit{ In the context of movie recommendation, what are some extrinsic factors that are relevant for decision-making?} \\
\textbf{ChatGPT}:\textit{ There are several extrinsic factors beyond the user's demographic information and viewing history that can be relevant for decision-making. These factors can include: Time of day, Weather, Events, Social media activity, Popular trends... For example, a user may be more interested in watching a romantic movie in the evening, and may prefer an action movie during the day.} \\
\textbf{User}:\textit{ On what basis do you propose that users may be more interested in watching a romantic movie in the evening? Please show statistics, with sources, if available.} \\
\textbf{ChatGPT}:\textit{ Upon further research, I couldn't find any empirical studies that provide conclusive evidence... However, I can provide some reasoning for why this may be the case: Mood... , Personal preferences... , Societal norms... } \\ 
\textbf{User}:\textit{ I see. So the reasoning behind " If it's raining outside, the user may be more interested in watching a movie that has a cozy, indoor setting." is also based on social norm, without any quantitative support?} \\
\textbf{ChatGPT}:\textit{ Yes, it is also based on social norms and common sense, rather than empirical evidence.}\\
\textbf{User}:\textit{ What type of movies would be recommend for a student on the eve of a critical examination?} \\
\textbf{ChatGPT}:\textit{ It may be advisable to recommend movies that are engaging but not too emotionally or mentally taxing. This will ensure that the student can relax and unwind without getting too distracted or overwhelmed before their exam.}\\
\textbf{User}:\textit{ What if a student is requesting movie recommendations on the eve of a critical exam, with a plan to travel afterward?} \\
\textbf{ChatGPT}:\textit{ In that case, the recommended movies should not only be engaging and relaxing but also align with their upcoming travel plans...}\\
\textbf{User}:\textit{ Would there be any potential concerns for suggesting moves that align with the student's upcoming travel plans, given it's the night before an important exam?} \\
\textbf{ChatGPT}:\textit{ There may be some potential concerns for suggesting movies that align with the user's upcoming travel plans, particularly if it's the night before an important exam. Here are a few potential considerations: 1. Anxiety: Watching a movie about the student's upcoming travel plans may increase their anxiety or excitement about the trip, making it difficult to focus on their studies... It may be advisable to recommend movies that are more relaxing and calming rather than ones that may increase the student's excitement or anxiety levels.}
\end{formal}

\section{Conclusion and Future work}

In this study, we establish ten fundamental principles that an AGR should adhere to and test ChatGPT's potential to serve as an AGR. While ChatGPT exhibits significant promise in fulfilling the principles, further research is required to address existing limitations. These limitations may include the model's inability to: proactively seek clarifications, differentiate between relevancy levels of various queries, retain metadata that is inconsistent with the most recent user input, discern causal relationships among user utterances, provide accurate information for less popular items, derive logically consistent insights from interaction history, identify embedded factual inconsistencies, and accurately utilize ratings in recommendation reasoning. A more comprehensive understanding of such limitations and the development of strategies to mitigate them will contribute to the refinement of ChatGPT's performance as an AGR.


\bibliographystyle{ACM-Reference-Format}
\bibliography{sample-base}

\end{document}